\documentstyle[amsfonts,epsfig,12pt]{article}

\def\mypagenumber{1}

\def\myend{\end{document}}

\def\Journal#1#2#3#4{{#1}{\bf #2} (#3) #4}

\def\CQG{\em Class.\ Quant.\ Grav.}
\def\NPB{{\em Nucl.\ Phys.} B}
\def\PLB{{\em Phys.\ Lett.} B}
\def\PRL{\em Phys.\ Rev.\ Lett. }

\def\PRD{{\em Phys.\ Rev.} D}

\def\CMP{\em Comm.\ Math.\ Phys. }

\normalsize

\newcounter{sxn}

\newcounter{axn}

\date{}

\newdimen\mybaselineskip
\newcommand{\beeq}{\begin{equation}}
\newcommand{\eneq}{\end{equation}}
\newcommand{\be}{\begin{eqnarray}}
\newcommand{\ee}{\end{eqnarray}}
\newcommand{\bpic}{\begin{picture}}
\newcommand{\epic}{\end{picture}}

\def\dd{\partial}
\def\la{\raise.16ex\hbox{$\langle$} \, }
\def\ra{\, \raise.16ex\hbox{$\rangle$} }
\def\go{\rightarrow}

\def\psibar{ \psi \kern-.65em\raise.6em\hbox{$-$} }
\def\mbar{ m \kern-.78em\raise.4em\hbox{$-$}\lower.4em\hbox{} }

\def\n@space{\nulldelimiterspace=0pt \mathsurround=0pt }
\def\huge#1{{\hbox{$\left#1\vbox to 20.5pt{}\right.\n@space$}}}

\def\myskip{\noalign{\kern 8pt}}
\def\myeqspace{\noalign{\kern 10pt}}

\def\boxit#1{$\vcenter{\hrule\hbox{\vrule\kern3pt
    \vbox{\kern3pt\hbox{#1}\kern3pt}\kern3pt\vrule}\hrule}$}
\def\bigbox#1{$\vcenter{\hrule\hbox{\vrule\kern5pt
     \vbox{\kern5pt\hbox{#1}\kern5pt}\kern5pt\vrule}\hrule}$}

\def\ignore#1{{}}

\begin{document}

\bibliographystyle{unsrt}
\footskip 1.0cm

\thispagestyle{empty}
\setcounter{page}{\mypagenumber}

             
\begin{flushright}{
OUTP-00-27-P\\
}

\end{flushright}

\vspace{2.5cm}
\begin{center}
{\LARGE \bf {Multi-instantons in ${\mathbb{R}}^4$ and   }}\\
\vskip 0.5 cm   
{\LARGE \bf { Minimal Surfaces in ${\mathbb{R}}^{2,1}$ }}

\vspace{2cm}
{\large  Bayram Tekin~\footnote{e-mail:~tekin@thphys.ox.ac.uk} }\\
\vspace{.5cm}
{\it Theoretical Physics, University of Oxford, 1 Keble Road, Oxford,
OX1 3NP, UK}\\  
\end{center}

\vspace*{1 cm}


\begin{abstract}
\baselineskip=18pt
It is known that self-duality equations for multi-instantons on a line in four dimensions are 
equivalent to minimal surface equations in three dimensional Minkowski space. 
We extend this equivalence 
beyond the equations of motion and show that topological number, instanton moduli space  and 
anti-self-dual solutions have representations in terms of minimal surfaces. The issue of topological
charge is quite subtle because the surfaces that appear are non-compact. 
This minimal surface/instanton correspondence allows us to define a metric on the configuration space of the gauge fields.
We obtain the minimal surface representation of an instanton with arbitrary charge. 
The trivial vacuum and the BPST instanton as minimal surfaces are worked out in detail. 
BPS monopoles and the geodesics are also discussed. 
\end{abstract}
\vfill



 
\newpage



\normalsize
\baselineskip=22pt plus 1pt minus 1pt
\parindent=25pt

\section{Introduction}
Over twenty years ago Comtet \cite{Comtet} showed that the equations for the 
cylindrically symmetric multi-instantons 
of Yang-Mills theory given by Witten \cite{Witten} correspond to minimal surface equations. 
He proved this equivalence at the level of the equations of motion.
I extend his analysis and show that multi instantons are represented by minimal surfaces in 
every aspects, including the topological charge, the moduli and the anti-self-dual solutions. 
In particular the issue of the topological charge of instantons in terms of
topological properties of the minimal surfaces is quite subtle because the minimal surfaces that 
appear are non-compact and have infinite total curvature. So there is no well-defined notion of 
Euler number for these surfaces. We will show that a finite (renormalized) topological
invariant, which will correspond to the topological charge of the instanton, 
can be defined for these surfaces. Through this construction there is also a natural way of 
defining a metric in the configuration space of the gauge fields.  

Minimal surfaces also show up in the self-dual solutions of 
Einstein's equations in four dimensions. 
Nutku \cite{Nutku, Aliev} demonstrated  that 
Gibbons-Hawking \cite{Gibbons2} multi gravitational 
instantons can be obtained from minimal surfaces in three dimensions.
The correspondence follows by showing that the equations for the 
Ricci-flat K\"{a}hler metrics with (anti) self-dual curvature are exactly
minimal surface equations in three dimensions. So the conclusion is that 
for every minimal surface there is a gravitational instanton.  

Immediate physical relevance of the minimal surface and instanton equivalence
is not clear. But one is tempted to speculate that 
this correspondence 
is reminiscent of string/gauge fields duality along the lines of 
\cite{Polyakov1, Polyakov2, Polyakov3}. According to Polyakov gauge invariant objects,
presumably Wilson loops, are expected to have string theory  representations.
In this paper we will show that certain multi-instantons (objects which have gauge invariant 
properties) are represented by minimal surfaces
which would mean a Euclidean version of gauge fields/strings duality.  
We will not pursue this interpretation any further in this paper but leave it to future work. 

The outline of the paper is as follows. In section 2 we give a review of the cylindrically symmetric
multi instanton solution in four dimensions. In section 3 we describe minimal surfaces in three 
dimensional Minkowski space and show that the equations describing the minimal surfaces are
equivalent to the  self-duality equations of the four dimensional gauge fields.
In section 3 we also define a metric on the configuration space of the gauge fields by using the
metric in the minimal surface. In section 4 we find the minimal surfaces that correspond to the
trivial vacuum solution and the BPST instanton and anti-instanton solutions. 
In section 5 by using dimensional reduction we describe charge one BPS monopole in terms of geodesics in the
minimal surfaces. Section 6 consists of conclusions and discussions.

\section{Multi-instantons }
In this section we will give a brief review of Witten's \cite{Witten} results. 
Multi-instantons on a line in ${\mathbb{R}}^4$ are finite action
solutions of the self-dual Yang-Mills fields which can be represented in the following form,
\be
&&A^{a}_{j}(\vec{x}) = {1\over r} \left[ \epsilon^a\,_{kj}\,
\hat{x}^k\,(1+\varphi_2)+ \delta^a\,_j\varphi_1 +(r A_1+\varphi_1)\hat{x}^a
\hat{x}_j\right]\nonumber \\
&&A^{a}_{0}(\vec{x}) = A_0 \hat{x}^a 
\ee
where all $\varphi_i$, $A_i$ are functions of the three dimensional radius $r\in [0,\infty]$ and the Euclidean 
time $t\in[-\infty,\infty]$.  
We will work exclusively with the gauge group $SU(2)$ so  $a = (1,2,3)$. $\hat{x}^a$ are unit vectors. 
Setting the coupling constant to unity, the
Euclidean Yang-Mills action reduces to the Abelian-Higgs model in a curved 
space.
\be
S_{YM}&=& {1\over{4}}\int_{{\mathbb{R}}^4} d^4x F^a_{\mu\nu}F^a_{\mu\nu}
\nonumber \\ 
&=& 8\pi\int_{U} d^2 x \, \sqrt{g} \Bigg\{ 
{1\over 2}g^{\mu\nu}D_\mu \varphi_i\,D_\nu \varphi_i 
+{1\over 8}g^{\mu\alpha}\,g^{\mu\beta}  F_{\mu\alpha} F_{\alpha \beta} +
{1\over {4}}(1 -\varphi_i^2)^2  \ \Bigg \}
\label{action1}
\ee
Where $F_{\mu \nu}= \dd_\mu A_\nu - \dd_\nu A_\mu$ and 
$D_\mu \varphi_i= \dd_\mu \varphi_i +\epsilon_{ij} A_\mu \varphi_j$. $U$ is the upper-half plane
with the Poincar\'{e} metric $g^{\mu\nu} = r^2\delta^{\mu\nu}$, yielding $ds^2= r^{-2}(dr^2 +dt^2)$.~\footnote{Clearly for these cylindrically symmetric fields Yang-Mills theory in ${\mathbb{R}}^4$ is equivalent to a 
non-conformal theory in $AdS_2 \times S^2$, where, $AdS_2$ has infinite and  
$S^2$  has a unit radius and the conformal structure is fixed by choosing the Poincar\'{e} 
metric on $AdS_2$.}   
The (Gaussian) scalar curvature is $K_U = -1$.  
From the self-duality (anti self-duality), $F^a_{\mu\nu}= \pm {\tilde{F}}^a_{\mu\nu}$  
condition equations of motion read as
\be
D_0\varphi_1 = \pm D_1\varphi_2 \hskip 1cm D_1\varphi_1 = \mp D_0\varphi_2 \hskip 1cm  
r^2 F_{01}= \pm(1-\varphi_1^2-\varphi_1^2 )
\label{self-duality}
\ee
The sign on the top refers the self-dual and the lower one to anti-self-dual solutions.
There is clearly  a $U(1)$ symmetry in this reduced theory.
For later use let us write down how this symmetry acts on the fields.
\be
\tilde{\varphi_1} = \varphi_1 \cos\theta +\varphi_2 \sin\theta,
\hskip 0.5cm  
\tilde{\varphi_2} = -\varphi_1 \sin\theta +\varphi_2 \cos\theta, 
\hskip 0.5cm 
\tilde{A_\mu} = A_\mu - \dd_\mu\theta
\label{symmetry}
\ee
$\theta$ is a function of $r$ and $t$. 
We can pick up the Lorentz gauge  $\dd_\mu\,A_\mu=0$ which can be solved 
by $A_\mu = \epsilon_{\mu\nu}\dd_\nu\psi$. Then defining 
$\varphi_1 = e^\psi\chi_1$ and  $\varphi_2 = e^\psi\chi_2$ the first two
equations in (\ref{self-duality}) reduce to the Cauchy-Riemann equations
for the analytic function, $f(z) = \chi_1 - i\chi_2$.     
\be
\dd_0\chi_1 = \dd_1\chi_2 \hskip 2 cm \dd_1\chi_1 = \dd_0\chi_2
\label{Cauchy-Riemann}
\ee
Where $z= r+it$. The last equation in (\ref{self-duality}) becomes the Liouville equation in the
curved space
\be
r^2 \Delta \psi = |f|^2e^{2\psi}-1
\ee
The most general solution to this equation is given by

\be
&&\psi = \log\Bigg({2r\over {(1 - |g|^2)|h|}}\Bigg), \hskip 2 cm 
g(z)= \prod_{i=1}^k \Bigg( {{a_i-z}\over {\stackrel{*}{a_i}+z }} \Bigg) 
 \nonumber \\
&&h(z)= -i\prod_{i=1}^k \big({\stackrel{*}{a_i}+z }\big)^2, \hskip 2.5 cm 
\varphi_1 -i\varphi_2 = h {dg\over dz}e^\psi 
\label{solution}
\ee
Where $|g|^2={\stackrel{*}{g}}$ $g$. 
For non-singular $\psi$ we have $|g|=1$ at $r=0$, $|g|<1$ for $r>0$ and $a_i$ are constants 
for which
$Re(a_i)>0$. $k=1$ solution corresponds to the vacuum and $k=2$ corresponds to
the BPST instanton. The locations of the zeros of $ {dg/dz}$ are gauge invariant and real part
of a zero of this function corresponds to the instanton size and the imaginary part
corresponds to the point on the $t$ axis where the instanton is located. A careful 
counting shows that 2 of the parameters in (\ref{solution}) are gauge artifacts and 
there are $2 (k-1)$ parameters for a generic solution  with $k-1$ topological number.
 
The topological charge of the theory is
\be
Q = \pm{1\over {8\pi^2}}\int\, d^4x  F^a_{\mu\nu}{\tilde{F}}^a_{\mu\nu}
= \pm {1\over {4\pi}}\int\,d^2x \epsilon_{\mu\nu}\,F_{\mu\nu}  
\ee
The four dimensional topological charge reduces to the magnetic flux in the 
Abelian-Higgs model. The magnetic flux is equal to the number of zeros of  ${dg/dz}$ multiplied by 
$2\pi$ giving $Q = k-1$   

For later use I will write down the topological charge in the following form 
\be
Q=\pm {1\over {2\pi}}\int_{-\infty}^{\infty}dt\,\int_0^{\infty}dr {1\over r^2}\,(1- \varphi_1^2- \varphi_2^2 )
\label{charge}
\ee

In the next section I will describe minimal surfaces in 
${\mathbb{R}}^{2,1}$ and show that they carry all the properties of the multi-instantons 
that we described above.

\section{Minimal Surfaces}

Before we give a detailed description of minimal surfaces let us mention that even a cursory 
look at the self-duality equations suggests that some kind of special surfaces will arise
in the configuration space of this theory. 
We start with four functions, $A_0, A_1, \varphi_1, \varphi_2$, of  $r$ 
and $t$.
The choice of gauge eliminates one of the functions, i.e. $A_0 = - A_1$, and we end up with three  which
describe a generic surface. Comtet \cite{Comtet} showed that self-duality equations 
(\ref{self-duality}) make this surface a minimal surface.

Let us recall \cite{Osserman} that a surface  
in ${\mathbb{R}}^{3}$ is a differentiable map $f$ from a domain $\Sigma$ into 
${\mathbb{R}}^{3}$. Such a surface, in the non-parametric form $z = f(x,y)$, 
will be minimal if it satisfies   
\be
f_{x\,x}\, (1+f_y^2) -2 f_x\, f_y\, f_{x\,y} + f_{y\,y}\, (1+f_x^2) =0,
\label{minsurface}
\ee
where $f_x$ denotes partial differentiation. This equation describes the surfaces with vanishing mean curvature
and it can be obtained by minimizing the area of the surface.
\be
A= \int_\Sigma \, \sqrt{ 1+ f_x^2 +f_y^2}\,dx\,dy 
\ee
A general conformal immersion solution to the minimal surface equation is given by the Weierstrass 
representation.~\footnote{There is a nice spinor representation of  minimal surfaces which may turn out to
be quite useful for physics \cite{Kusner}. The spinor bundle $S$ over $\Sigma$ is a two dimensional 
vector bundle 
which is $S= \Lambda^0\oplus\Lambda^{(1,0)}$. So one defines the spinor $\xi= (g,\eta)$. 
The minimal surface equation is given by the Dirac equation for this spinor, $D\,(\xi)= 0$}
   
\be
f= {\mbox{Re}}\,\Bigg( \int (1- g^2, i(1+g^2), 2g)\eta \Bigg):\Sigma \rightarrow   {\mathbb{R}}^{3} 
\label{Weierstrass}
\ee 
Where  $g$ is a holomorphic function and $\eta$ is a holomorphic 1-form.

For our purposes non-parametric description, which would mean eliminating $r$ and $t$ from the 
gauge fields, 
is not suitable. Moreover
not all the minimal surfaces can be represented in the non-parametric form. So we will use 
an other description which will make the instanton-minimal surface 
correspondence more transparent. As it is clear from the previous section the minimal surfaces
that will appear are smooth sub-domains of the upper-half-plane ,$U$, with the Poincar\'{e} 
metric of constant negative curvature. Due to a theorem of Hilbert \cite{Thurston} we can not
embed $U$ in ${\mathbb{R}}^{3}$. So we will consider the Minkowski space ${\mathbb{R}}^{2,1}$ 
as the embedding space.     

Using the notations of \cite{Flanders} let us
consider a surface $\Sigma$ in ${\mathbb{R}}^{2,1}$. Define orthonormal frames, 
$(\vec{e_\mu})$, on the surface. Here $\mu$ = $(1,2,3)$. They satisfy  
$\vec{e_\mu}\cdot\vec{e_\nu}= g_{\mu\nu}$. Where 
$g_{\mu\nu} = \mbox{diag}(1,1,-1)$. 
And we take $(\vec{e_3})$ to be orthogonal to the surface.
We expect our surface to be smooth (at least twice differentiable) so we can use isothermal
coordinates $(u,v)$ for which the line element on the surface takes the form 
\be
 ds^2 = \lambda^2(u,v)\, ( du^2 + dv^2)
\ee
There are two 1-forms on the surface which we define as 
\be
\sigma_1= \lambda(u,v) du , \hskip 2 cm \sigma_2= \lambda(u,v) dv
\label{1forms}
\ee
A point, $\vec{x}$, which is restricted to move on the surface satisfies
\be
d\vec{x} = \sigma_i \vec{e_i},
\ee  
where summation is implied for $i =(1,2)$. We need to write down  how the 
frame moves. Let us define
\be
d \vec{e_\mu}= \omega_{\mu\nu}\vec{e_\alpha}\,g^{\nu\alpha}, 
\ee 
where  $\omega_{\mu\nu}$ are antisymmetric one forms. Using these 
{{\it{structure equations}}
one can derive the {\it{integrability conditions}} 
(or Gauss-Codazzi equations). 
\be
d\omega_{\mu\nu}-
\omega_{\mu\alpha}\wedge\omega_{\beta\nu}\,g^{\alpha\beta}= 0 \hskip 1 cm 
d\sigma_\mu - \omega_{\mu\nu}\wedge \sigma_\alpha \,g^{\nu\alpha}=0  
\ee
For the surface we have $\sigma_3 =0$. These formulae define a general
surface which is not necessarily minimal. Gaussian and the mean curvature
of this surface are defined in the following way  
\be
K(u,v)= -{1\over\lambda^2}\Delta \log\lambda, 
\hskip 1 cm  
{1\over 4}
\epsilon_{\mu\nu\alpha}\,\omega_{\beta\gamma}\wedge\sigma_{\delta}g^{\mu\beta}
g^{\nu\gamma} g^{\alpha\delta}  
\equiv H(u,v) \sigma_1 \wedge \sigma_2   
\ee
Here $\sigma_1 \wedge \sigma_2$ is the area of the surface.  
For the minimal surfaces $H(u,v) =0$.  
So eventually  we have the following sets of equations for the minimal surface
\be
&&d\sigma_1 = {\tilde{\omega}}\wedge\sigma_2 ,\hskip 1 cm 
d\sigma_2 =   -{\tilde{\omega}}\wedge\sigma_1, \hskip 1 cm 
d\omega_1  =  {\tilde{\omega}}\wedge\omega_2 \hskip 1 cm \nonumber \\
&&d\omega_2  =- {\tilde{\omega}}\wedge \omega_1, \hskip 1 cm 
d{\tilde{\omega}}  = \omega_1\wedge\omega_2 \hskip 1 cm
\sigma_1\wedge\omega_1 +\sigma_2\wedge\omega_2 = 0 \nonumber \\
&&\sigma_1\wedge\omega_2 -\sigma_2\wedge\omega_1 = 0
\ee
We have defined $\omega_{1\,2} = {\tilde{\omega}} $ ,  
$\omega_{1\,3} =-\omega_1 $ 
and   $\omega_{2\,3} =-\omega_2$. These one-forms can be expressed as linear
combinations of $\sigma_1$ and  $\sigma_2$. Looking at the above equations
and using (\ref{1forms}) one obtains 
\be
&&\omega_1 = p(u,v)\, d u + q(u,v)\, d v \nonumber \\ 
&&\omega_2 = q(u,v)\, d u - p(u,v)\, d v  \nonumber \\
&& {\tilde{\omega}} = a(u,v)\, d u  + b(u,v)\, d v
\ee

Finally we have the following differential equations which 
describe the minimal surfaces
\be
\dot{p} - q' = a\,p + b\,q 
\hskip 1 cm  \dot{a} - b' =  p^2+q^2 \hskip 1cm 
\dot{q} + p' = a\,q - b\,p 
\label{minimal}
\ee
In addition to these there are two more equations which will
give us the conformal factor in the metric.
\be
\dot{\lambda} = -a\lambda 
\hskip 2 cm  \lambda' = b\lambda 
\label{conformalfactor}
\ee
where $\dot{p} = {\dd p\over \dd v}$ and $p' = {\dd p\over \dd u}$ etc...
The claim is that these equations are equivalent to the self-duality equations (\ref{self-duality})
this can be proved by making the following identifications~\footnote{ Equivalently one can 
choose the following identification; $v=r$, $u=t$, $q= {\varphi_1\over r}$, 
$p= {\varphi_2\over r}$, $a= A_0 -{1\over r}$ and $b= A_1$.} 

\be
&&u = r, \hskip 1 cm v = t, \hskip 1cm  b = {1\over r} - A_o  \nonumber \\
&&p = {\varphi_1\over r}, \hskip 1cm  q = {\varphi_2\over r} \hskip 1cm a = -A_1
\label{identifications1}
\ee

Anti-self-dual solutions can be obtained by 
\be
 b = {1\over r} + A_o   \hskip 0.5cm a = A_1 \hskip 0.5cm
q = {\varphi_1\over r}, \hskip 0.5 cm  p = {\varphi_2\over r} 
\label{identifications2}
\ee

With these identifications, recasting Comtet's \cite{Comtet} argument,  
we have shown that the minimal surface 
equations (\ref{minimal}) are exactly
equivalent to multi-instanton equations (\ref{self-duality}).  
We also need to find a topological invariant in the minimal surface which should 
correspond to the  topological charge of the instanton.
The Gaussian curvature of the surface can be calculated to give
\be
K(r,t)=  {1\over r^2 \lambda^2} (\varphi_1^2 + \varphi_2^2).
\label{Gauss}
\ee
A naive topological invariant for the minimal surface would be the total curvature,
\be
\chi =\int dA\, K(r,t)= \int_{-\infty}^{\infty}dt\,\int_0^{\infty}dr {1\over r^2}\,(\varphi_1^2 + \varphi_2^2 )
\label{charge2}
\ee
which clearly diverges. We will resolve this issue in the next section.
 
Minimal surface equations enjoy the $U(1)$ symmetry (\ref{symmetry}) in the following way
\be
&&\tilde{p} = p \cos\theta +q \sin\theta,
\hskip 1 cm  
\tilde{q} = -p \sin\theta +q \cos\theta, 
\nonumber \\
&&\tilde{a} = a + \theta' \hskip 3cm  \tilde{b} = b + \dot{\theta}
\label{symmetry2}
\ee
The gauge invariance of the equations involving $\lambda$ can also be shown
by first noting that, 
\be
\lambda(r,t)= {\mbox{exp}}\{ -\int_{t_0}^t a dt + \int_{r_0}^r b dr \}  
\ee
We know how $a$ and $b$ transform from (\ref{symmetry2}). So we find that
$\lambda$ transforms as
\be
\tilde{\lambda(r,t)} = 
\lambda(r,t){\mbox{exp}} \{ -\int_{t_0}^t\theta' dt + 
\int_{r_0}^r \dot{\theta} dr \}   
\ee
Using this one can show that (\ref{conformalfactor}) are gauge invariant
if the gauge parameter $\theta$ is a harmonic function.

We have shown that self-duality equations are in one to one correspondence
with the equations that define a minimal surface. We have two more equations
(\ref{conformalfactor}) on the minimal surface which will give us the 
metric on the surface. We will interpret this metric as a metric on the
configuration space of the gauge fields. Using the equations 
(\ref{identifications1}) and the solution (\ref{solution}) one obtains the 
conformal factor and the metric for the self-dual solutions as
\be
\lambda(r,t) = r e^{-\psi(r,t)} = (1- |g|^2)|h|, \hskip 1cm 
ds^2 = (1- |g|^2)^2|h|^2 (dr^2 + dt^2)    
\label{metricselfdual}
\ee
This is the metric on a minimal surface ( and configuration space of gauge fields) that corresponds to
a charge $k-1$ instanton. All the details of the minimal surfaces can be read off directly from the
solutions of the self-duality equations. Of course the other way around is also possible by solving the
minimal surface equations. For the anti-self-dual solutions we have
\be
\lambda(r,t) = r e^{\psi(r,t)} = {r^2\over (1- |g|^2)|h|}, \hskip 1cm 
ds^2 = {r^4\over (1- |g|^2)^2|h|^2} (dr^2 + dt^2)    
\label{metricantiselfdual}
\ee
In $\lambda(r,t)$ I have suppressed an overall constant factor.

\section{BPST as a minimal surface}

We start with the trivial vacuum solution, $k =1$, which is both self-dual and 
anti-self-dual. The general solution (\ref{solution}) yields $\varphi_2= -1$ and 
$\varphi_1 = A_0= A_1=0$. The metric on the field space (the minimal surface) 
 and the Gaussian curvature are,
\be
ds^2_{vacuum}= r^2( dr^2 + dt^2) \hskip 2 cm K_{vacuum} = {1\over r^4} 
\ee
This is the Robertson-Walker metric. There is a singularity at 
the origin and the horizon is at infinity. 
From gauge theory side we know that the instanton number of the
trivial vacuum is zero. As a minimal surface we need to look at the  
total Gaussian curvature which turns out to be infinite. This is not a big
surprise as we saw in the previous section. Our minimal surfaces are embedded in the upper-half plane
with the Poincare metric, for which $K_U = -1$. The total curvature of the $U$-plane 
is infinite simply because it is non-compact and has an infinite area.  
The lesson we learn from the vacuum solution is that we need to renormalize 
the total curvature of the minimal surface in order to get the correct 
topological charge of the gauge theory solutions. Let us write down a general formula 
which will be true for all instanton solutions. ~\footnote{This 
renormalization/regularization is rather standard in gravity 
and it was first outlined by 
Gibbons and Hawking \cite{Gibbons} 
in the context of four dimensional gravity. See also \cite{Hawking}.  
The reader might wonder why the boundary 
contributions to the topological charges are not being considered.  
The reason is that the boundary term would be an integral of the trace of the
second fundamental form of the surface,  which is the
the mean curvature which vanishes for the minimal surfaces by definition.} 
\be
Q =  {1\over 2 \pi}\, \Bigg(\int K_U dA + \int K_{\Sigma}dA \Bigg)    
\label{topcharge}
\ee
Using the Gaussian curvature $K_{\Sigma}$ (\ref{Gauss}) of the minimal 
surface and $K_U= -1$ of the Poincare plane one gets the topological number for the instanton (\ref{charge}).
This is a perfectly a well-defined finite number which, as we stated before, is equal to $\pm (k-1)$    

Now we will find the minimal surface corresponding to the 
BPST \cite{Belavin} instanton.
Choosing $k=2$ one has
\be
h(z) = -i \big({\stackrel{*}{a_1}+z }\big)^2 \,
\big({\stackrel{*}{a_2}+z }\big)^2 \hskip 1 cm 
g(z)=   {(a_1-z)\,(a_2-z)\over{ ({\stackrel{*}{a_1}+z})\,
         ({\stackrel{*}{a_2}}+z )}}
\ee
There are four arbitrary parameters but a check of the zero of   $dg(z)/dz$ 
shows that two of 
these parameters are redundant. 
The physical parameters are
\be
t_0 = {{\mbox{Im}} (a_1\,a_2) \over{{\mbox{Re}}(a_1 + a_2)}}, \hskip 2 cm 
\rho^2 = -t_0^2 + {{\mbox{Re}}[{\stackrel{*}{a_1}}{\stackrel{*}{a_2}}(a_1 +a_2)]
\over{\mbox{Re}}(a_1+a_2)},   
\ee  
where $t_0$ is the location of the instanton on the time axis and the
$\rho_0$ is the size of it.
The metric for the self-dual solutions follows as
\be
ds^2_{BPST} = r^2[r^2 +(t-t_0)^2 + \rho^2 ]^2 \, ( dr^2 + dt^2)    
\label{BPSTmetric}
\ee
It is crucial that we get the $r^2$ factor in front. It is the factor that
cancels the infinity in the topological charge of the minimal 
surface ( instanton ) as we have seen in the vacuum case. This is again 
a Robertson-Walker type metric.   
Given the metric above one readily calculates the
curvature of the minimal surface that corresponds to the BPST instanton.
The curvature has a singularity at the origin.
To obtain the topological charge  we use (\ref{topcharge}). The second
factor in (\ref{BPSTmetric}) gives a $2\pi$ contribution to the total curvature and we obtain
$Q= 1$.
For the sake of completeness let us denote that
\be
\psi(r,t)= - \log{1\over 2}[r^2 +(t-t_0)^2 + \rho^2]
\ee  
The rest of the functions (for the gauge-fields or the minimal surfaces ) 
can be found trivially. Anti-BPST solution with topological charge $-1$, 
follows similarly.
\be
ds^2_{anti-BPST}=  {r^2\over [r^2 +(t-t_0)^2 + \rho^2 ]^2} \, ( dr^2 + dt^2)   \ee

\section{BPS Monopole and Geodesics}
It is quite instructive to apply the results of the previous sections to the BPS monopoles.
In pure Yang-Mills theory BPS monopoles can appear in a number of different ways. For example
Rossi \cite{Rossi} showed that infinite number of instantons ( $k \go \infty$ ) with the same 
instanton size and regular separation (periodic instantons) on a line appear as a BPS monopole 
in the limit of vanishing separation.
A more direct (at the end equivalent) way is to consider all the 
fields to be independent of ``time' $t$.~\footnote{We can only get a charge one monopole through this
construction. Hitchin \cite{Hitchin} gave an implicit construction of multi-monopoles from geodesics.}
So self-duality equations become
\be
A_0\varphi_2 = \dd_r\varphi_2 - A_1\varphi_1 ,\hskip 0.5cm   
\dd_r\varphi_1 + A_1\varphi_2 =A_0\varphi_1, \hskip 0.5 cm
-r^2 \dd_r A_0 = 1-\varphi_1^2 - \varphi_2^2
\label{monopole}
\ee 
The solution given by \cite{Prasad} is 
\be
A_0 ={1\over r} - \coth (r), \hskip 1cm \varphi_2 = {r\over \sinh(r)}, \hskip 1 cm 
A_1= \varphi_1=0 
\ee
In fact one obtains a one-dimensional moduli of solutions corresponding to the value of $A_0 (\infty)$,
which I have assumed to be $-1$ here. 

After dimensional reduction the minimal surface equations become
\be
q' = -bq, \hskip 0.5cm b'= -(p^2+q^2), \hskip 0.5 cm  p' = -bp, \hskip 0.5 cm \lambda' = b \lambda 
\ee
The solution follows as
\be
a = p= 0, \hskip 1 cm q = {1\over\sinh(r)}, \hskip 1 cm b= \coth(r) \hskip 1 cm \lambda= \sinh(r)
\ee
This is a geodesic in the minimal surface. The curvature of this geodesic is
\be
K(r) = {1\over \sinh^4(r)}
\ee
Along the lines of the discussion of topological charge from the previous section we need to renormalize
the total curvature of this geodesic to get the correct topological number for the BPS monopole. 
The geodesics in the upper half plane
are given by semi-circles and the vertical lines that are orthogonal to the t-axis. In the limit of
vanishing $t$ only the vertical geodesic at the origin survive, $ds^2= r^{-2}dr^2$ . Its curvature is
$-1$. So we need to subtract the total curvature of this  vertical geodesic from the BPS geodesic.
It follows that~\footnote{To take care of the trivial factor of $2\pi$ in front of (\ref{charge2})
one can think that we compactify the t-direction on a circle of unit radius}
\be
Q_{BPS} = -\int_0^\infty dr\,K(r)\,\lambda(r)^2 +   \int_0^\infty dr\,{1\over r^2}= 1   
\ee

\section{Conclusion}

Extending the analysis of Comtet \cite{Comtet} we have shown that cylindrically symmetric
multi-instantons are equivalent to minimal surfaces in three dimensional Minkowski space.
At the level of the equations of motion this equivalence follows rather directly. 
The issue of topological charges was subtle and we showed that the ``Euler number'' of the minimal
surface requires a renormalization to get the correct topological number for the corresponding 
instanton. One can also interpret this renormalization as adding an Einstein-Hilbert action (Euler number)   
to the Abelian-Higgs model action that was discussed in the first section. Gravity in two dimensions 
is non-dynamical and we know that for our particular model we have $AdS_2$ with the Poincar\'{e} metric.

These instantons with topological charge $n$  have $2n$ dimensional moduli spaces. 
The corresponding minimal surfaces  have $2n$ moduli as it can be seen from the most general 
explicit solution of the minimal surface equations. The dimension of the moduli space  
don't have a simple expression in terms
of the genus of the surface since our surfaces are not closed. 
Anti-self dual solutions are related to the self-dual solutions in a rather non-trivial way as 
can be seen from equations (\ref{metricselfdual}) and (\ref{metricantiselfdual}).

Through our construction there is a natural metric defined on the configuration space of the gauge 
fields which is the metric on the minimal surface. We have given two explicit examples of these. 
The first one is the trivial vacuum solution and the the other one is the BPST instanton.
We worked out the details of the minimal surfaces that correspond to these solutions.
 
Charge one BPS monopole/geodesics equivalence is a natural byproduct of our construction after
dimensional reduction. In this case one again needs to renormalize the total curvature of the
BPS geodesic to get the correct topological charge for the BPS monopole.

Gibbons-Hawking gravitational multi-instantons can also derived be from minimal surfaces as it was shown
by Nutku \cite{Nutku}. The issue of the topological charge and the moduli 
in this correspondence  needs to be studied in detail.  

A possible further direction of research would be to understand how this correspondence fits 
in the picture of gauge fields and string/duality. Self-dual gauge fields is 
 perhaps a simple system  of gauge fields where one can establish with some rigor an equivalence  
between field theory and string theory. We leave these discussion for  future work.

\section{Acknowledgments}
I would like thank G. Dunne, A. Kovner, A. Lukas and M. Schvellinger for
useful discussions. 
This work was supported by  PPARC Grant PPA/G/O/1998/00567.

\vskip 1cm

\leftline{\bf References}  

\renewenvironment{thebibliography}[1]
        {\begin{list}{[$\,$\arabic{enumi}$\,$]}  
        {\usecounter{enumi}\setlength{\parsep}{0pt}
         \setlength{\itemsep}{0pt}  \renewcommand{\baselinestretch}{1.2}
         \settowidth
        {\labelwidth}{#1 ~ ~}\sloppy}}{\end{list}}

\myend